\documentclass[prl,twocolumn,tightenlines,showpacs,amsmath]{revtex4}
\usepackage{epsfig} 

\begin{document} 

\title{ Relativistic electron-ion recombination 
in the presence of an intense laser field } 
                                      
\author{ C.M\"uller, A.B.Voitkiv and B.Najjari} 
\affiliation{ Max-Planck-Institut f\"ur Kernphysik, 
Saupfercheckweg 1, D-69117 Heidelberg, Germany } 

\date{\today} 

\begin{abstract} 

Radiative recombination of a relativistic electron 
with a highly charged ion in the presence of an intense 
laser field is considered. 
Various relativistic effects, caused by 
the high energy of the incoming electron and its strong coupling 
to the intense laser field, 
are found to clearly manifest themselves 
in the spectra of the emitted $\gamma$-photons.  

\end{abstract}

%\pacs{34.10+x, 34.50.Rk, 34.80.Qb}
%\pacs{42.50.Hz, 25.75.Dw, 32.80.-t}
\maketitle 

%PACS:
%34.80.Qb Laser-modified scattering (of electrons and positrons)
%34.80.Lx Recombination, attachment, and positronium formation 
%41.75.Ht Relativistic electron and positron beams 
%32.80.Wr Other multiphoton processes (photoionization/excitation)
%34.50.Rk Laser-modified scattering and reactions (of atoms/molecules)
%52.20.-j Elementary processes in plasmas
%52.20.Fs Electron collisions

% body of paper begins here 

Radiative recombination (RR) of a free electron with 
an ion is a fundamental process which plays an
important role in all kinds of astrophysical and laboratory plasmas. It
represents the inverse of the photoeffect and has been studied for a
long time \cite{Hahn}. 

More recently, the availability of intense laser
sources has triggered a growing interest in RR in the presence of an
external laser field which can lead to strong modifications of the
field-free properties of the process. When, for example, the laser
photon frequency is resonant with the electron transition energy, the
field stimulates the RR process and substantially enhances its rate.
Corresponding experimental investigations rely on merged beams of ions,
electrons, and photons in a storage ring \cite{Schwalm}. Very recently,
multiphoton-assisted RR of low-energy electrons into barium ions in a
microwave cavity has been observed in a weak-coupling regime
\cite{Gallagher}.

Modern powerful laser devices are capable of producing field intensities
in excess of $10^{20}$ W/cm$^2$ in the optical and near-infrared
frequency domain. Electrons exposed to such strong fields may be
accelerated to highly relativistic energies, for example, through
laser-plasma interactions \cite{Malka}. Modifications of many atomic
collision processes such as Mott and M{\o}ller scattering, or Bethe-Heitler
pair creation due to the presence of a relativistically strong
background laser field have been theoretically studied 
during the last decade \cite{ehl-new}-\cite{bete-heitler}. 

The theoretical investigation of laser-assisted radiative 
recombination (LARR), however, has so far been restricted 
to the nonrelativistic regime where the dipole-approximation 
to the field applies (see
\cite{Kuchiev,kam-ehl,Milosevic,Bivona,phase} and references therein).
The main emphasis was placed upon the $X$-ray energy spectrum. 
Effects from the intensity profile of a focused laser pulse have also been
addressed \cite{Milosevic} and the possibility 
of phase control has been shown \cite{phase}. 
The laser field may also affect the $X$-ray polarization \cite{Bivona}. 

Note also that LARR represents the last step in the coherent process 
of high-harmonic generation from gas targets
which is intensively being studied both theoretically and experimentally
for 20 years. In this process, RR of field-ionized atomic electrons 
occurs through laser-driven recollisions, with the highest harmonic
photon energies achieved of about 1\,keV \cite{Seres}. The relativistic 
domain of high-harmonic generation, however, is inaccessible until
now since at optical laser intensities exceeding $\sim 10^{17}$
W/cm$^2$ the magnetic-field induced Lorentz drift motion of the electron
prevents efficient recollisions \cite{Salamin,Klaiber}. This circumstance
provides additional motivation for studies of relativistic
LARR utilizing free electron and ion beams.

In this letter we extend the consideration of LARR into the relativistic
domain where various characteristic modifications of the process are
demonstrated to occur. Relativistic effects arise from the high initial
electron energy (in the MeV range) and the large laser 
field strength-to-frequency ratio leading to strong electron-field coupling.   
In particular, a very large number of low-frequency 
laser photons participate in the process whose  
total energy and momentum are imprinted 
on the emitted high-frequency ($\gamma$-ray) photon 
influencing its energy and angular distributions 
in a distinctive non-dipole manner. 

The laser field is assumed to be a classical monochromatic 
plane wave of circular polarization 
which is switched on and off adiabatically 
at $t \to -\infty$ and $t \to +\infty$, respectively. 
We shall describe this field by the vector potential 
taken in the form  
\begin{eqnarray} 
{\bf A}({\bf r},t) = a_0 
\left( {\bf e}_1 \cos \varphi  
+ {\bf e}_2 \sin \varphi \right); \,      
\varphi  =  \omega_0 t - {\bf k}_0 \cdot {\bf r}, 
\label{laser-field}  
\end{eqnarray} 
where ${\bf r}$ and $t$ are the space-time coordinates, 
$\omega_0$ and ${\bf k}_0$ are the frequency and wave vector, 
${\bf e}_1$ and ${\bf e}_2$ are 
the polarization vectors (${\bf e}_i \cdot {\bf e}_j = \delta_{ij}$, 
${\bf e}_i \cdot {\bf k}_0 =0$) and 
$a_0 = c F_0/\omega_0$ is 
the amplitude of the vector potential with 
$F_0$ and $c$ being the strength 
of the laser field and the speed of light, 
respectively.   

The transition amplitude for the radiative 
recombination reads \cite{at-un}   
\begin{eqnarray}
S_{fi} = -i \int_{-\infty}^{+\infty} dt % 
\langle \Psi_f(t) |\hat{W} |\Psi_i(t)\rangle,    % 
\label{ampl-gen}  
\end{eqnarray}   
where $ | \Psi_i(t) \rangle $ and $ | \Psi_f(t) \rangle$ 
are the initial and final states 
of the system 'electron+radiation field' 
and $\hat{W}$ is the interaction with the radiation field. 
The latter is chosen in the form 
%\begin{eqnarray}
$\hat{W} = \mbox{\boldmath $\alpha$} \cdot \hat{ {\bf A}}_{\gamma} $ 
%\label{int-rad}
%\end{eqnarray}
where $\mbox{\boldmath $\alpha$} = (\alpha_x, \alpha_y, \alpha_z)$
are the Dirac matrices and 
\begin{eqnarray}
\hat{ {\bf A}}_{\gamma}({\bf r},t)&=& \sum_{ {\bf k}, \rho } %  
\sqrt{ \frac{ 2\pi c^2 }{ V \omega_k }} 
{\bf e}_{{\bf k}, \rho } % 
\left( c^{+}_{ {\bf k}, \rho  } % 
e^{i( \omega_k t - {\bf k} \cdot {\bf r} )}  %
+ C.C. \right).    % 
\label{potent-rad}
\end{eqnarray} 
In (\ref{potent-rad}) 
$ c^{+}_{ {\bf k} \rho }$ is  
the creation operator for a photon with momentum 
${\bf k}$, polarization vectors   
$ {\bf e}_{ {\bf k} \rho } $ ($\rho=1,2$)  
and frequency $\omega_k= c |{\bf k}|$, 
$V$ is the normalization volume 
for the radiation field and 
the sum runs over all photon modes. 

The states $\Psi_i(t)$ and $\Psi_f(t)$ 
are given by  
\begin{eqnarray} 
| \Psi_i(t) \rangle = \psi_i(t) \, | 0 \rangle, \, \,  
| \Psi_f(t) \rangle = \psi_f(t) \, | {\bf k} \rho \rangle,    
\label{states-1}  
\end{eqnarray} 
where $ | 0 \rangle $ and $| {\bf k} \rho \rangle $ 
denote states of the radiation field with 
no photons and with one photon having momentum 
${\bf k}$ and polarization $\rho$. The initial and 
final states of the electron, $\psi_i$ and $\psi_f$, 
are solutions of the Dirac equation 
\begin{eqnarray} 
i \frac{\partial \psi}{\partial t } = 
c \mbox{\boldmath $\alpha$} \cdot % 
\left( \hat{\bf p} + \frac{1}{c} {\bf A} \right) % 
\psi + V_0 \psi + \beta c^2 \psi,      
\label{dirac}  
\end{eqnarray} 
where $V_0 = - Z/r$ is the interaction 
between the electron and the ionic nucleus 
and $\beta$ is the Dirac matrix. 
Both in the initial and final states 
the electron is subject to 
the simultaneous presence of two fields: 
the field of the nucleus 
and the laser field. Since an exact solution 
of such a problem is not known suitable approximations 
to describe these states are needed.  

The incident electron is supposed to have 
relativistic asymptotic momentum $p_i$ and energy $E_i$.  
Assuming that the charge $Z$ of the nucleus 
satisfies the condition $Z \ll c$ 
one can neglect the Coulomb interaction 
in the initial state.  
At the same time the low-frequency laser field 
can very substantially affect the motion 
of the incident electron, including 
its momentum and energy, and  
the interaction of the electron 
with this field should be taken 
into account to all orders. Therefore, 
we shall approximate the initial state of the 
electron using the Gordon-Volkov solution \cite{Gordon} 
for a free electron moving in the electromagnetic field 
described by the vector potential (\ref{laser-field}).   

Let us now assume that 
although the laser field is 
quite strong, its strength $F_0$ 
nevertheless remains much smaller 
than the typical nuclear field 
$F_a \sim Z^3$ acting on the electron 
bound in the ground state of the ion \cite{ff1}. 
Then the motion of the electron in the final state 
will be practically determined only by 
its interaction with the nuclear charge 
while the influence of the laser field 
on the tightly bound 
electron is comparably very 
weak and can be neglected.   

Note, however, that in the gauge 
in which the potential of the laser field 
is given by Eq.(\ref{laser-field}), one 
cannot simply approximate $\psi_f$ by 
$\phi_0({\bf r}) \, \exp(-i \varepsilon_0 t)$, 
where $\phi_0$ is the wave function of the 
electron in the ground state of the ion and 
$\varepsilon_0$ is the corresponding 
electron energy. Indeed, it is not 
difficult to see that in this gauge the  
term in the wave equation (\ref{dirac}), 
which describes the interaction between 
the electron and the laser field, is 
of the order of $F_0 \, Z/\omega_0$  
and it may be not small compared 
to the term $|V_0| \sim Z^2$, 
even if the condition $F_0 \ll Z^3$ 
is extremely well fulfilled. 

Therefore, in order to obtain an 
appropriate approximation 
for the final state of the electron 
we following \cite{kam-ehl}, \cite{abv} 
first set $\psi_f = \exp \left( -i \frac{ {\bf A} \cdot {\bf r}}{c} \right) \chi$.  
Inserting this ansatz into Eq.(\ref{dirac}) 
one can easily convince oneself that 
the term representing the interaction 
between the electron and the laser field 
in the obtained Dirac equation for the function $\chi$ 
is now of the order of $F_0 /Z $. 
Thus, now the electron-laser field interaction term 
is much smaller than $|V_0| \sim Z^2$. Therefore,  
we may replace 
$\chi$ by $\phi_0({\bf r}) \, \exp(-i \varepsilon_0 t)$ 
and, thus,  
\begin{eqnarray} 
\psi_f = \phi_0({\bf r}) \, \exp(-i \varepsilon_0 t) % 
\exp \left( -i \frac{ {\bf A} \cdot {\bf r}}{c} \right).  
\label{final-state} 
\end{eqnarray}

Inserting the Gordon-Volkov state and (\ref{final-state}) 
into (\ref{ampl-gen}) and assuming that the 
condition $ v_{i\perp}/\omega_0 \gg r_0 $  
is fulfilled (where $v_{i\perp}$ is the part of   
the initial electron velocity $v_i$ perpendicular 
to ${\bf k}_0$ and $r_0 \sim 1/Z$ is the size of the final bound state 
of the electron) we obtain 
\begin{eqnarray}
S_{fi} &=& -i 2 \pi \frac{2 \pi c^2}{V \omega_k} \, %  
\, \sqrt{ \frac{mc^2}{E_i} } 
\nonumber \\ 
&& \times \sum_{n=-\infty}^{n=+\infty} G_n \, 
\delta\left(n \omega_0 + \varepsilon_0 + \omega_k - \xi \omega_0 - E_i \right).   
\label{ampl}  
\end{eqnarray}   
In the above expression 
\begin{eqnarray} 
G_n &=& \int d^3 {\bf r} % 
\exp(i ({\bf p}_i - {\bf k} - n {\bf k}_0 + \xi {\bf k}_0) \cdot {\bf r}) 
\nonumber \\ 
&& \left( {\bf e}_{ {\bf k} \rho } \cdot {\bf f} % 
\, J_n(\Lambda) + \right. 
\nonumber \\   
&& \left. \exp(i \chi_{\bf p}) \, \exp( i {\bf b} \cdot {\bf r} ) % 
\, {\bf e}_{ {\bf k} \rho } \cdot {\bf f}_{+} % 
\, J_{n-1}(\Lambda) + \right. % 
\nonumber \\ 
&& \left. \exp(-i \chi_{\bf p}) \, % 
\exp( - i {\bf b} \cdot {\bf r} ) \, % 
{\bf e}_{ {\bf k} \rho } \cdot {\bf f}_{-}  % 
\, J_{n+1}(\Lambda) \right),  
\label{G}
\end{eqnarray} 
where ${\bf f}= \phi_0^{\dagger} \, \mbox{\boldmath $\alpha$} \, u({\bf p}_i,s)$,  
${\bf f}_{\pm} = \phi_0^{\dagger} \, \mbox{\boldmath $\alpha$} \, 
 \eta_{\pm}\, u({\bf p}_i,s)$ with $u({\bf p},s)$ being 
the free Dirac spinor for an electron with momentum ${\bf p}$ and spin $s$, 
\begin{eqnarray} 
\eta_{\pm} = \frac{a_0}{4 c } \, \frac{ \omega_0/c +  % 
\mbox{\boldmath $\alpha$} \cdot {\bf k}_0} % 
{\omega_0 E_i/c^2 - {\bf k}_0 \cdot {\bf p}_i} 
\, \mbox{\boldmath $\alpha$} \cdot ({\bf e}_1 \mp i {\bf e}_2 ),    
\label{eta}
\end{eqnarray} 
$J_n$ is the Bessel function with argument  
\begin{eqnarray} 
\Lambda &=& \lambda - {\bf d} \cdot {\bf r}; \, \,  
{\bf d} = \frac{a_0^2}{\lambda \, c^2} \, 
\frac{ ({\bf e}_1  \cdot {\bf p}_i) \, {\bf e}_2 - 
({\bf e}_2  \cdot {\bf p}_i) \, {\bf e}_1 } 
{\omega_0 E_i/c^2 - {\bf k}_0 \cdot {\bf p}_i}   % 
\nonumber \\ 
\lambda &=&  %  
\frac{a_0/c}{ \omega_0 E_i/c^2 - {\bf k}_0 \cdot {\bf p}_i } 
\sqrt{( {\bf e}_1  \cdot {\bf p}_i )^2 + 
( {\bf e}_2  \cdot {\bf p}_i )^2} % 
\label{lambda} 
\end{eqnarray} 
and 
\begin{eqnarray}
\xi &=&  \frac{a_0^2 }{2 (\omega_0 E_i - c^2 {\bf k}_0 \cdot {\bf p}_i) }; \, \,  
\chi_{\bf p} = \arctan \left( \frac{{\bf e}_2  \cdot {\bf p}_i } % 
{{\bf e}_1  \cdot {\bf p}_i   } \right) 
\nonumber \\ 
{\bf b} &=& \frac{a_0^2}{\lambda^2 \, c^2} \, 
\frac{ ({\bf e}_1  \cdot {\bf p}_i) \, {\bf e}_1 +  
({\bf e}_2  \cdot {\bf p}_i) \, {\bf e}_2 } 
{\omega_0 E_i/c^2 - {\bf k}_0 \cdot {\bf p}_i}.  % 
\label{ksi} 
\end{eqnarray} 
In what follows we, based on Eqs. (\ref{ampl})-(\ref{ksi}),  
consider the reaction $e^-$ + Sn$^{50+}$ $\to$ 
Sn$^{49+}$(1s) + $\gamma$ 
occurring in a laser field 
with $F_0=7.5$ a.u. ($I \approx 4 \times 10^{18}$ W/cm$^2$), 
$\omega_0=0.055$ a.u. ($1.5$ eV), ${\bf e}_1=(1,0,0)$ 
and ${\bf e}_1=(0,1,0)$ in which 
the initial electron momentum $p_i=10 mc$ 
directed along the $x$-axis.  
\begin{figure}[t]  
\vspace{-0.25cm}
\begin{center}
\includegraphics[width=0.37\textwidth]{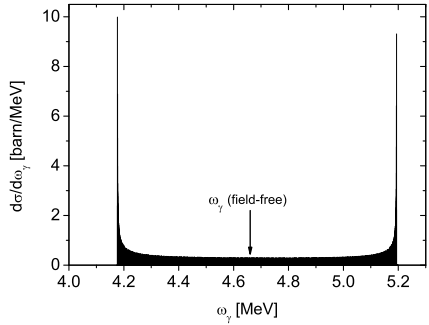}%{el-laser1.eps}
\end{center}
\vspace{-0.75cm} 
\caption{ \footnotesize{ 
Energy spectrum of the $\gamma$-ray photons emitted 
in LARR of a relativistic electron ($p_i=10mc$) with 
a bare Sn nucleus ($Z=50$) in a circularly 
polarized laser field with $F_0=7.5$ a.u. 
and $\omega_0 =0.055$ a.u.  crossing 
the beam under $90^0$. } } 
\label{figure1}
\end{figure}

Figure \ref{figure1} shows the energy spectrum 
of the $\gamma$-ray photons. The spectrum consists of a multitude 
of lines each separated by a laser photon energy, thus 
forming a quasi-continuous distribution with a total 
width of $\Delta\omega_{\gamma} \sim 1$\,MeV and very pronounced 
side wings. Without the laser field, the spectrum would 
comprise a single line at 
$\omega_{\gamma}=E_i-\varepsilon_0\approx 4.66$ MeV. 

The motion of an electron in the presence of a strong 
laser field has quasi-classical character. Therefore, 
similarly as in the nonrelativistic case (see e.g. \cite{kam-ehl}), 
certain features of the relativistic LARR energy spectrum 
can be understood in classical terms: In the presence 
of the field the instantaneous electron energy is modulated,
\begin{eqnarray} 
E(\varphi) &=& E_i + U_p + \lambda\omega_0\cos\varphi 
\nonumber \\ 
&=& E_i + \frac{c^2 F_0^2}{2 \omega_0^2 E_i} 
+ \frac{c p_i}{E_i} \frac{c F_0}{\omega_0} \cos\varphi 
\label{E}
\end{eqnarray}
so that the available kinetic energy depends on 
the instant phase of recombination. 
While the ponderomotive energy 
$U_p = \xi \omega_0 \approx 25$\,keV 
leads to a relatively small shift of the center of the spectrum, 
the oscillating term in Eq.\,(\ref{E}) causes a spectral 
width of $\Delta\omega_{\gamma}=2\lambda\omega_0\sim 1$\,MeV 
between the minimum and maximum energies corresponding 
to recombination occurring at $\cos \varphi= -1 $ 
and $\cos \varphi= 1$, respectively. 

In the quantum picture the broad spectrum results 
from the emission or absorption of $n$ laser 
photons during the RR process, with the boundaries 
$|n|\lesssim n_{max} \simeq \lambda$ determined by the properties 
of the Bessel functions. While already the classical 
consideration gives the range of the energy spectrum, 
of course only a quantum consideration 
can predict the shape of the spectrum.   

It is known that in the nonrelativistic domain 
the spectral width grows with the kinetic energy of 
the incoming electron (\cite{Kuchiev}-\cite{kam-ehl}). 
In the present case, however, we observe 
that the width becomes practically 
independent of the incoming electron energy, 
i.e. it saturates.  This occurs because 
$\lambda\propto v_i$ [see Eq.\,(\ref{lambda})] and 
the electron velocity $v_i$  
cannot exceed the speed of light. 
This implies in fact that the width and 
shape of the spectrum in Fig.~\ref{figure1}  
(where $v_i\approx c$) are 'universal' in the sense 
that they will remain practically unchanged 
when $E_i$ is further increased. 
Enhancement of the electron 
energy would shift the center of the spectrum 
to correspondingly higher energies but, 
even in the limit $E_i\to\infty$, the number of 
laser photons participating in the process would 
not increase noticeably.  

Note also that the ratio of the spectral 
intensity at the side wings as compared with 
the plateau region in between scales with 
$\lambda^{1/3}$ and amounts to $\approx 30$ 
here which is significantly larger than 
in the nonrelativistic regime. 
In particular, each side wing contributes 
$1.5\%$ to the total LARR cross section when 
a narrow energy window of $1$ keV 
is taken into account.

\begin{figure}[t]  
\vspace{-0.25cm}
\begin{center}
\includegraphics[width=0.38\textwidth]{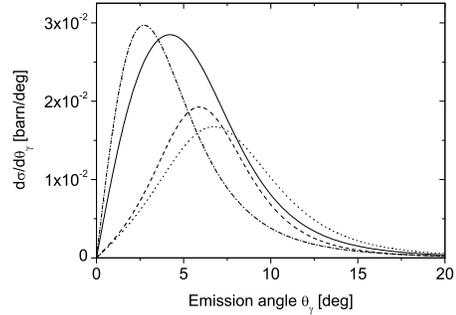}
\end{center}
\vspace{-0.75cm} 
\caption{ \footnotesize{Angular distributions 
of the emitted $\gamma$-ray photons with respect 
to the incoming electron momentum. 
The dash-dotted curve shows the field-free case, 
whereas the solid line refers to LARR with 
the laser parameters of Fig.~\ref{figure1}.
The dotted and dashed lines show the spectra 
for the energy side wings around $\omega_{\gamma} \approx 4.2$\,MeV 
and $\omega_{\gamma} \approx 5.2$\,MeV (see Fig.~\ref{figure1}), 
respectively, within an energy interval of $\delta\omega_{\gamma} = 1$\,keV; 
the latter two curves are multiplied by a factor of 30.}
} 
\label{figure2}
\end{figure}

Characteristic relativistic signatures also 
arise in the angular emission spectrum, 
shown in Fig.~\ref{figure2}. 
In the nonrelativistic domain the photons produced 
by RR into the ground state are emitted 
mainly in the direction perpendicular 
to the momentum of the incident electron, 
forming a typical dipole pattern.
In contrast to that 
the $\gamma$-rays are emitted 
essentially along the direction of the incoming electron.  
This feature is also known from relativistic  
field-free RR \cite{Eichler}. However, 
as compared with the field-free case, 
the LARR spectrum in Fig.~\ref{figure2} 
is shifted to larger angles and broadened. 

The shift is most pronounced 
in the energetic side wings where 
it is almost solely caused by 
the momentum carried by the laser photons 
(and thus would be absent in the nonrelativistic domain) 
and where it may again be understood in classical 
terms by inspecting the instantaneous electron momentum 
${\bf p}(\varphi)={\bf p}_i + \frac{1}{c}{\bf A}(\varphi)+[c{\bf p}_i\cdot{\bf A}(\varphi)+\frac{1}{2}{\bf A}^2]\frac{{\bf k}_0}{\omega_0 E_i}$, 
with the components
\begin{eqnarray}
p_x(\varphi) &=& p_i +\frac{F_0}{\omega_0}\cos\varphi\,;\ \ p_y(\varphi)=\frac{F_0}{\omega_0}\sin\varphi\,; \nonumber\\
p_z(\varphi) &=& \frac{U_p}{c} + 
\frac{p_i c}{E_i} \frac{F_0}{\omega_0}\cos\varphi.  
\label{p}
\end{eqnarray}
The $x$ and $y$ components lie in the polarization 
plane of the field and show a corresponding modulation 
due to the coupling to the electric component of the field, 
whereas the electron momentum along the $z$ axis arises 
solely from the light pressure exerted by the field momentum. 
By Eq.\,(\ref{p}), the typical polar angle $\theta_{\gamma}$ 
under which the electron impinges on the ion 
before recombination is given by $\tan\theta_{\gamma}=p_\perp/p_x$ 
with $p_\perp=\sqrt{p_y^2+p_z^2}$ 
and depends on the phase $\varphi$. 
For the minimum instantaneous electron energies 
($\varphi=\pi$) we have $p_\perp = p_z \approx F_0/\omega_0 \approx mc$, 
$p_x \approx 9mc$ and $\theta_{\gamma} \approx 6.3^\circ$, 
whereas for $\varphi=0$ where the maximum electron 
energy occurs we obtain $p_x \approx 11mc$, 
$p_\perp \approx mc$ and $\theta_{\gamma}\approx 5.2^\circ$. 
Both $\theta$ values are in agreement 
with the side-wing distributions shown in Fig.~\ref{figure2}.  

A similar, but somewhat more complicated, 
explanation can also be found for the position 
of the maxima of medium-energy $\gamma$-rays. 
Let us consider, 
for example, the center of the energy plateau in 
Fig.~\ref{figure1} where $\cos\varphi=0$ and thus 
a net number of $n=0$ laser photons participate in the process. 
The corresponding angular spectrum (not shown) peaks 
at small angles around $\theta\approx 2.5^\circ$  
which coincides with the field-free case 
(see Fig.~\ref{figure2}). This can be understood 
by noting that now, within a single 
laser cycle, there are two possible phases 
$\varphi_+=\pi/2$ and $\varphi_-=3\pi/2$, 
both with $\cos\varphi_\pm=0$ but opposite 
directions of $p_y(\varphi_\pm) \approx \pm mc$,  
whose contributions to the LARR amplitude interfere. 
The peak for $n=0$ may thus be considered 
as arising from the 'average' electron momentum 
with $\frac{1}{2}[p_y(\varphi_+)+p_y(\varphi_-)]=0$ 
and $p_x=p_i$. 

For increasing photon numbers $n$, 
with $0\ll|n|\ll n_{\rm max}$, 
the interference effect still exists 
but is becoming less pronounced. 
When the side wings are eventually reached 
($n\approx\pm n_{\rm max}$), 
both quantum paths merge into a single one 
and the classical picture becomes applicable. 
The peak of the total LARR angular spectrum, 
which represents a sum over the contributions 
from all laser photon numbers, hence lies 
in between the field-free peak and the side-wing peaks.

Finally we note that the presence of the laser field 
not only influences the $\gamma$-ray spectra but also 
has an impact on the total cross section increasing 
the latter in the case under consideration 
by about a factor of $2$ compared 
to the laser field-free case.  

In conclusion, we have considered radiative recombination 
of a relativistic electron with a highly charged ion  
assisted by an intense laser field of circular 
polarization. We have shown that the field 
substantially modify the shape 
of the spectra of the emitted $\gamma$ photons.  
We have discussed relativistic effects 
caused by the high energy of the incoming electron 
and its strong coupling to the laser field 
resulting in very large energy-momentum exchanges 
between the recombining electron-ion system and 
the field. 
 
An observation of the predicted effects
in principle is feasible in many high-power laser laboratories worldwide
because such lasers can be used to generate the relativistic electron
beams and highly charged ions required \cite{Malka,Salamin}.
%Alternatively, the latter may be produced by a dedicated ion source
%\cite{EBIT}. 
A particularly suited facility where relativistic LARR can be explored 
is the GSI in Darmstadt, Germany. 
%It is specialized in heavy-ion research, including RR studies, and since
%last year also operates the petawatt laser system PHELIX \cite{gsi}.

This work was supported in part by the Alliance Program of the Helmholtz
Association (HA216/EMMI).


\begin{thebibliography}{99}
\bibitem{at-un} Atomic units are used throughout 
unless otherwise indicated.
\bibitem{ff1} Provided $Z \gg 1$,   
the condition $F_0 \ll F_a$ 
is not very restrictive. 
For instance, for $Z=50$ 
one has $F_a \sim 10^5$ a.u. 
which means that even fields produced by 
the most powerful modern lasers 
($F_0 \stackrel{<}{\sim} 3 \times 10^2$ a.u.)  
are still very well fit 
into this condition.
\end{thebibliography}
\end{document}